\documentclass[prb,twocolumn,showpacs]{revtex4}
\usepackage{amsmath,bm,graphicx}

\begin{document}

\author{M.~Letz$^{a}$}
\author{W.~Mannstadt$^{a}$}
\author{M.~Brinkmann$^{a}$}
\author{E.~M\"orsen$^{b}$}

\affiliation{$^{a}$ Schott Glas, Research and Development, D-55014~Mainz, German
y}
\affiliation{$^{b}$ Schott Lithotecc AG, Otto-Schott Str. 13,
D-07705 Jena, Germany}

\title{Spatial dispersion in CaF$_2$ caused by the vicinity
of an excitonic bound state.} 

%>>>> The author is responsible for formatting the 
%  author list and their institutions.  Use  \skiplinehalf 
%  to separate author list from addresses and between each address.
%  The correspondence between each author and his/her address
%  can be indicated with a superscript in italics, 
%  which is easily obtained with \supit{}.

%\authorinfo{
%Send corrrespondence to M.Letz, e-mail: Martin.Letz@schott.com, phone 
%++49~6131~66~7201}
%% NB: when using amstex, you need to use @@ instead of @

%%%%%%%%%%%%%%%%%%%%%%%%%%%%%%%%%%%%%%%%%%%%%%%%%%%%%%%%%%%%% 
%>>>> uncomment following for page numbers
\pagestyle{plain}    
%>>>> uncomment following to start page numbering at 301 
%\setcounter{page}{301} 
 
%  \begin{document} 

%%%%%%%%%%%%%%%%%%%%%%%%%%%%%%%%%%%%%%%%%%%%%%%%%%%%%%%%%%%%% 
\begin{abstract}
The microscopic mechanism beyond the optical anisotropy of an ionic
crystal which occurs for short wavelengths is investigated. The
electron-hole, two particle propagator and its analytical behaviour
close to the band edge of the one particle continuum plays a major
role for the mechanism of this optical anisotropy. Especially for an
ionic crystal the two particle bound state, the exciton, is of special
importance. 
In this way we argue that the so called ``intrinsic
birefringence'' in CaF$_2$ is neither intrinsic to the material nor it
is birefringence. Instead it is spatial dispersion caused by the
vicinity of a dispersive optical absorption given by the excitonic
bound state. We propose a model which connects the bound state
dispersion with the band structure and a model potential for a
screened coulomb interaction. Based on these considerations we predict
a wavelength dependence of the dielectric function approaching close to the
bound state level $\epsilon \sim (\lambda - \lambda_0)^{-1}$, where
$\lambda_0$ is the wavelength of the excitonic bound state level. 
\end{abstract}

%>>>> Include a list of keywords after the abstract 


\maketitle

%%%%%%%%%%%%%%%%%%%%%%%%%%%%%%%%%%%%%%%%%%%%%%%%%%%%%%%%%%%%%
\section{INTRODUCTION}
\label{sect:intro}  % \label{} allows reference to this section

There is no doubt that CaF$_2$ is the key component for establishing
157nm UV lithography. Its perfect cubic symmetry together with a
great homogeneity of large single crystals which are very
stable under normal conditions makes it a perfect material for
homogeneous optical lens design.

However, there is a severe shortcoming of the
material on which the attention was focused within the last year. This
is the so called ``intrinsic birefringence''.\cite{burnett2001} For
the first glance it 
is astonishing. Looking at standard textbooks (e.g
\cite{nye79}) we note that a cubic crystal does not show any
birefringence. 
On the other hand experiments observe a spatial dependence of the dielectric properties.
Light propagating
along different directions through the crystal shows a 
deviation between different polarization directions which is of
the order of 10$^{-6}$
nm/cm for UV light with 157nm. 
Or with other words the dielectric
function depends on the 
direction in which the light propagates through the crystal and shows 
a difference which is of the order of 10$^{-6}$.

How can this obvious misfit between standard textbooks and recent
observations in CaF$_2$ be resolved? The answer leads back to a
physical phenomenon which is known as spatial dispersion.\cite{landaulifshitz8} 

The statement ``a cubic system does never show birefringence''
can be proven exactly, since in a cubic system the ellipsoid
corresponding to the dielectric function is always a sphere.
The question remains: What is a ``system''? The crystal of CaF$_2$ has
exactly cubic symmetry but the ``system'' of light interacting with
this perfectly cubic crystal shows a deviation from cubic symmetry. 
This is due to the fact that the light carries a ${\bf q}$ vector
pointing in the vacuum in the direction of propagation.
This is a
consideration which has been already made more than one century ago
\cite{lorentz1875}. However the speed of light is much larger than any
speed in a solid state system and therefore one can usually safely
neglect the wave-vector dependence of the propagating light. This
statement is identical to the wavelength of light being in the
optical window much larger than any length scale in a solid.

There are however two important exceptions where the assumption made
above ceases to be valid. The first (I) is when one is by far leaving the
optical window. This is the case at X-ray scattering experiments. Here
the wavelength of light is of the order of the lattice constant and
this causes a wave-vector (spatial) dependence of
the resulting scattering signal. The second case (II) which is valid
for CaF$_2$ and many other solids is the case of light in the vicinity 
of a dispersive
absorption
occurring also at small wavevectors (large length scales).
Since the dielectric function is an analytic function it
has to follow causality which is expressed in the Kramers-Kronig
relation between its real and imaginary part. The vicinity of a band
edge can cause spatial dispersion in a semiconductor system
(see e.g. \cite{cardona71}). 

In a semiconductor, the absorption is mainly dominated by the
band edge and an exciton, the two particle bound state of the
electron--hole pair only plays a minor role. The situation changes if
we look at ionic crystals like CaF$_2$. Here screening of the bare
coulomb interaction is less effective which expresses itself in deep
excitonic bound states. This is expressed in the exciton intensity
being inverse proportional to the third power of the static dielectric
function $I \sim \epsilon_0^{-3}$.\cite{phillips70} The connection
between a dispersive excitonic bound state and spatial dispersion has
been already discussed for KI. \cite{meseguer84}

When discussing possible alternative material designs to CaF$_2$ it is
important to understand the microscopic mechanism beyond the spatial
dispersion in great detail and it is in
particular important to know and predict the exciton dispersion. 

The paper is organized as follows: In section \ref{section1} we
define the system in the case when the approximation of locality
breaks down. In section \ref{section2} we 
review the connection between the electronic part of the dielectric
function and the bare two-particle propagator. We
look at the
free two particle propagator in the absence of any screened coulomb
attraction and its connection to the band structure. 
The connection between the band structure, which can e.g. be based on
DFT (density functional theory) calculations, and the dynamic
polarization is investigated in  
section \ref{section3}. In the
following (section \ref{section4}) we discuss analyticity based on the
resulting pole structure of the dielectric function. 
This describes to a large part the situation in a semiconductor where
no deep bound state is present.
Section (\ref{section5}) serves to introduce the two body problem of
the electron hole 
pair which leads to the excitonic bound state when a reasonable
assumption for the weakly screened coulomb attraction has been made. 
Based on the model assumptions made, we discuss the spatial dispersion
of CaF$_2$ close to the bound state level in section \ref{section6}
and compare our predictions with the experimental work
of.\cite{burnett2001}  

\section{System of CaF$_2$ interacting with light}
\label{section1}

Light has a ${\bf q}$ vector pointing along its
propagation direction. 
This wave-vector becomes important when the light is interacting with a
long wavelength (${\bf q}\rightarrow 0$) excitation of the system.
In an isotropic medium this wave vector dependence
will only depend on the absolute value of ${\bf q}$. In liquids and
glasses this leads e.g. to the observation of hydrodynamic modes
(phonon modes) in light scattering experiments.\cite{latz01} 
In single crystals the situation is different as not only the
dependence on the absolute value of the wavevector of the light is
important but also the direction 
of the wave vector with respect to the crystal orientations.
The crystal of CaF$_2$ possesses ideal cubic symmetry. That means in the
static limit that
also all response functions have to show this symmetry. Dynamically
there are situations especially close to a strong excitation at long
wavelengths, 
when the breaking of cubic symmetry due to the incident light gets
important. Therefore not only the dynamic of a dielectric response has
to be considered but also the wave-vector dependence. 
The electric field ${\bf E}$ is connected with the electric induction
${\bf D}$ in a non-local way.
\begin{equation}
{\bf D}({\bf q},\omega) = \epsilon({\bf q},\omega) {\bf E}({\bf q},\omega)
\end{equation}
where $\epsilon({\bf q},\omega)$ is the dynamic dielectric
function. For the physics of the spatial dispersion in CaF$_2$ we will
focus on the part of the dynamic dielectric function which is caused
by electronic excitations.
In general the dielectric function is an analytic function defined on
the complex $\omega$-plane. Along the real axis it has an real and
imaginary part. 
\begin{equation}
\epsilon({\bf q},\omega) = \epsilon'({\bf q},\omega) + i
\epsilon''({\bf q},\omega) 
\end{equation}
Only in the static limit $\omega \longrightarrow 0$
one can assign a physical value to the real and imaginary part
separately. The real part is connected with polarizabilities and the
imaginary part with mobile charges ($e''(\omega) = \lim_{\omega \rightarrow
0} 4 \pi \sigma(\omega)/\omega$) which are absent in an insulator. For
a general crystal the dielectric function is a tensor. Since we want
to keep the equations as transparent as possible we do not write
tensor indices. We assume a coordinate system placed along the main
axes of the cubic crystal. For a cubic
crystal like CaF$_2$ we can concentrate on specific directions of the
${\bf q}$ vector. Symmetry considerations allow to decompose all
tensorial quantities into its symmetry components. This will provide
rules to fill all the components of the
tensorial dielectric function.

   \begin{figure}
   \begin{center}
   \begin{tabular}{c}
\\[-2cm]
   \includegraphics[height=7 cm]{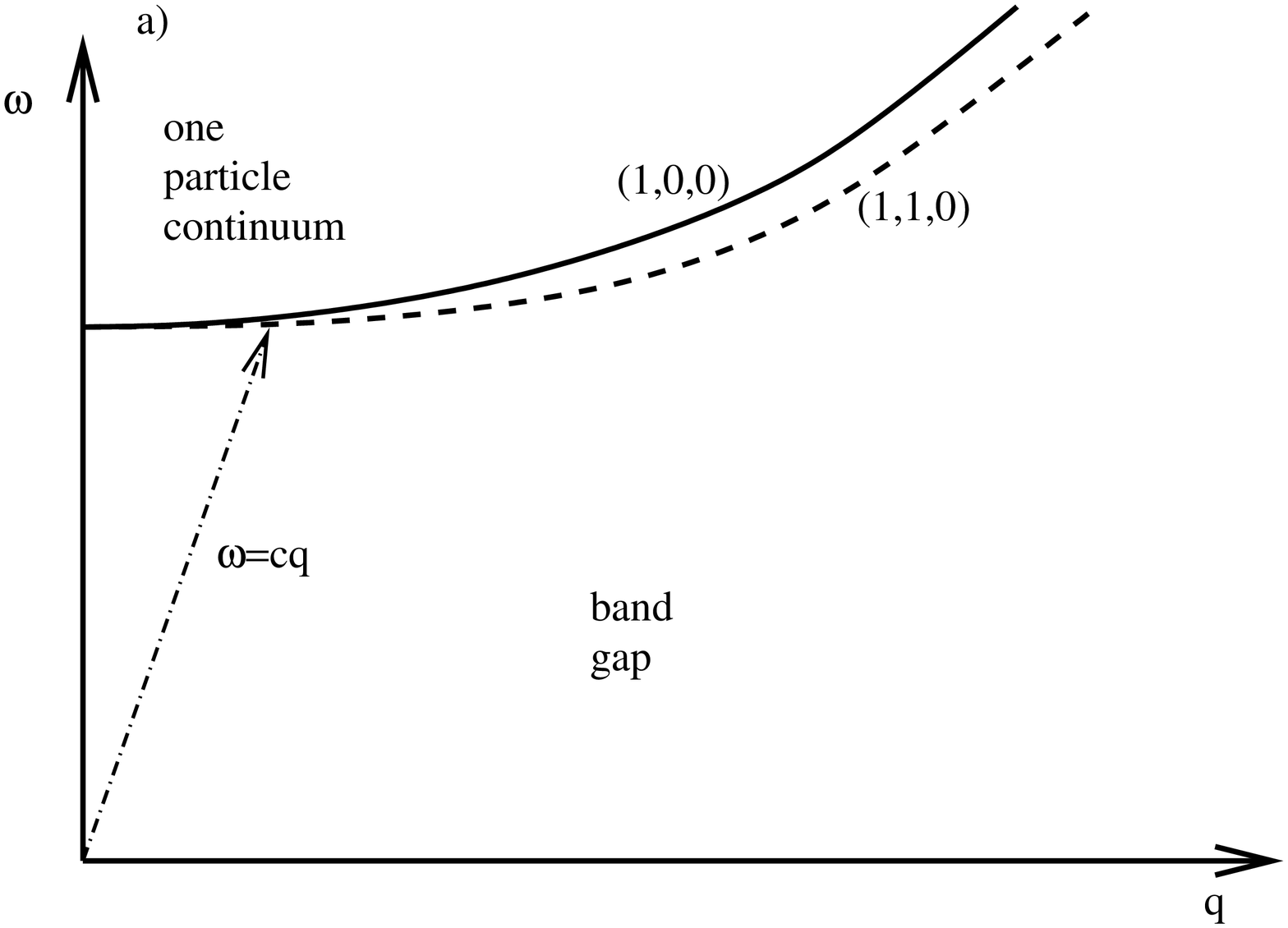} \\
   \includegraphics[height=4cm]{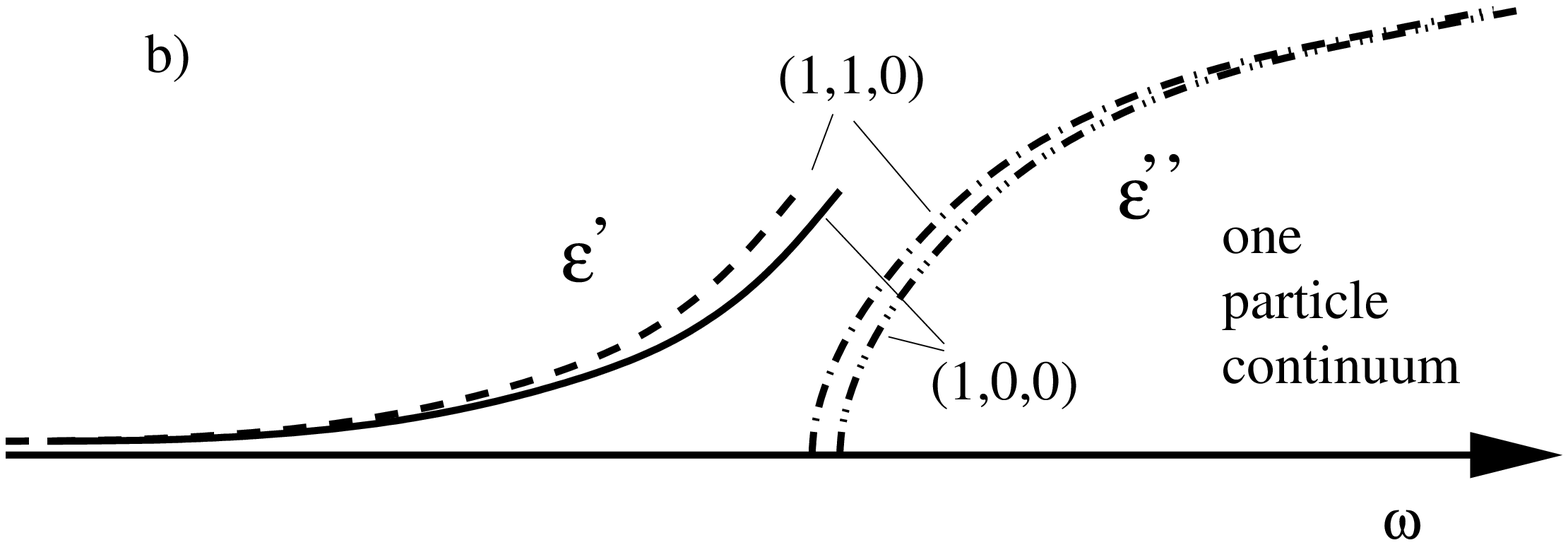}
   \end{tabular}
   \end{center}
   \caption[example] 
   { \label{fig:fig1} 
Schematic plot of the pole structure
of the two particle correlation function $\Pi^0({\bf q},\omega)$
 as it can occur for different
   directions of the wave vector. 
$\Pi^0({\bf q},\omega)$ is in the case when no excitonic bound state
   exists proportional to the dielectric function.
In Fig. a) we have plotted the band
   gap. The band gap can have a different shape when looking along
   different ${\bf q}$ directions. Above the band gap the two particle
   continuum starts where there is always a non zero imaginary part of
   the dielectric function. The light dispersion $\omega=c q$ is
   plotted schematically. In Fig. b) the real and imaginary part of
   the dielectric function resulting from Kramers-Kronig relations
   are plotted in a schematic way. When approaching the band gap the
   real part of the dielectric function bends up (but will not diverge
   in a 3D system) and the imaginary part will show a
   $\sqrt{\omega-\omega_0}$ behaviour. 
}
   \end{figure}

\section{The dynamic dielectric function}
\label{section2}

The optical frequency regime is above the infrared optical phonon
modes and below excitations over the band gap. In this regime the
dynamical dielectric function can be decomposed in two parts.
\begin{equation}
\epsilon({\bf q},\omega) = \epsilon_{\infty} + \epsilon_{e^--h^+}({\bf
q},\omega) 
\end{equation}
Where $\epsilon_{\infty}$ is the IR-dielectric function which in a
cubic crystal like CaF$_2$ will be not only real but completely
isotropic. The subscript $\infty$ refers to frequencies above the IR
optical phonon modes but is still far below exciatations over the bandgap.
Wave-vector dependence only enters from the second part which
stems from electron hole excitations. The electron-hole dielectric
function is proportional to the proper electron-hole polarization propagator
\cite{fetwal71}
\begin{equation}
\epsilon_{e^--h^+} ({\bf q},\omega) \sim \Pi^*({\bf q},\omega)
\end{equation}
Where $\Pi^*({\bf q},\omega)$ is the propagator of an electron hole
pair with total momentum ${\bf q}$. In the following we do not want to
focus on details but rather want to outline the pole structure of this
two particle propagator since it gives further insight into the
problem.

\section{Connection to the band structure}
\label{section3}

If we look for the first glance at the bare electron hole propagator
$\Pi^0({\bf q},\omega)$ we can make the connection to the band
structure. 
\begin{equation}
\label{eq:p0}
\Pi^0({\bf q},\omega) = \frac{1}{N} \sum_{\bf k} 
\frac{f({\bf k},{\bf q},\omega)}{\omega - (E^{e^-}({\bf k}+{\bf q}) -
E^{h^+}({\bf k}))}
\end{equation}
The detailed form of the numerator can be taken from textbooks
\cite{fetwal71} while the pole structure is connected with the
electronic band structure. In general Eq. (\ref{eq:p0}) should also
contain a sum over all bands. Since we are approaching the band gap
from the low frequency site and we know that CaF$_2$ has a direct band gap
we are for a first glance only interested
in the highest valence band $E^{h^+}({\bf k})$ and the
lowest conduction band $E^{e^-}({\bf k})$ around the $\Gamma$ point at
${\bf k}=(0,0,0)$. Even in a cubic crystal the band structure will show a
different ${\bf k}$ dependence when looking along different directions
in the crystal. In particular the $(1,0,0)$ direction will differ from
the $(1,1,0)$ and $(1,1,1)$ direction.

   \begin{figure}
   \begin{center}
   \begin{tabular}{c}
\\[-2cm]
   \includegraphics[height=7 cm]{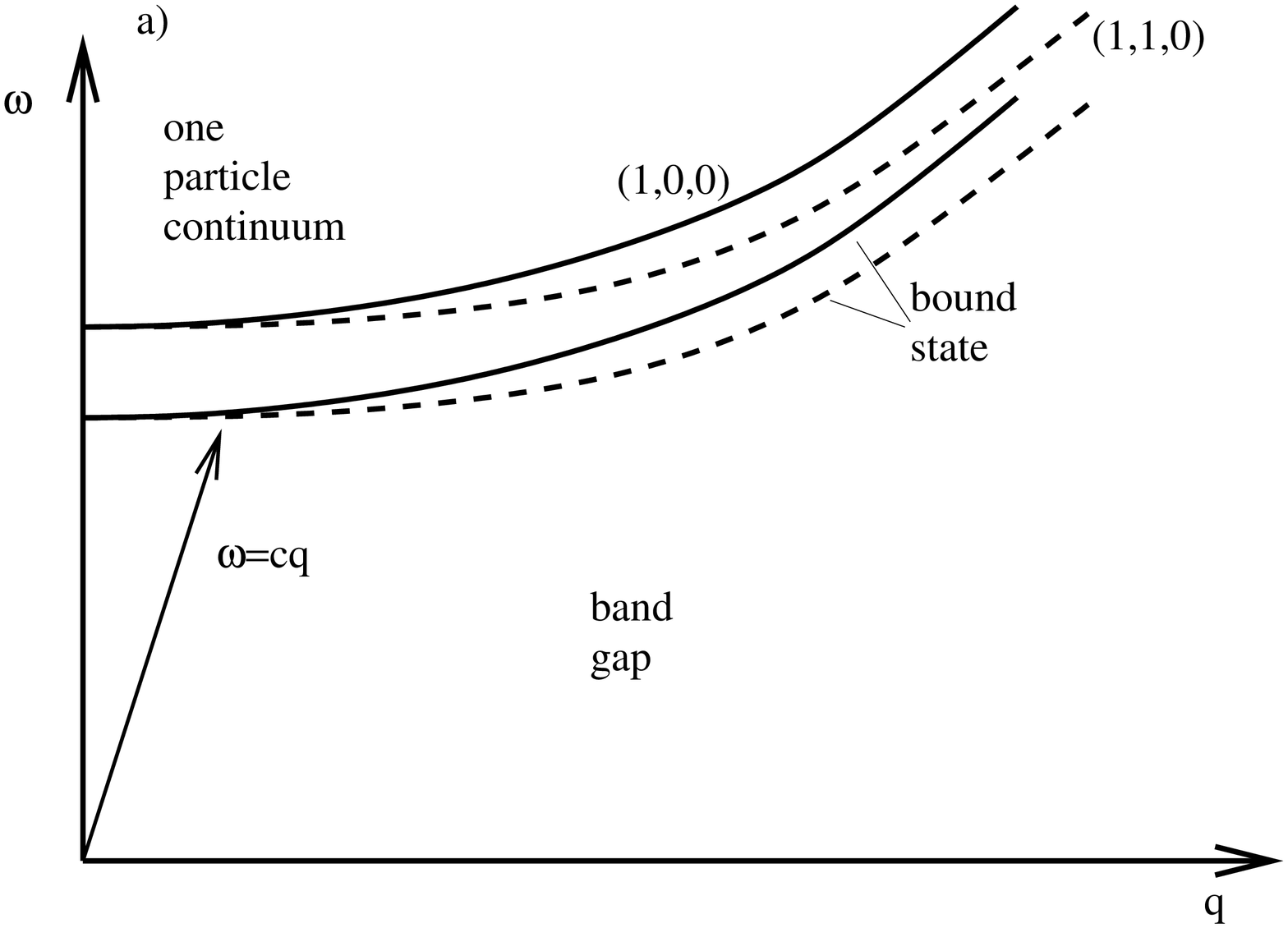} \\
   \includegraphics[height=4cm]{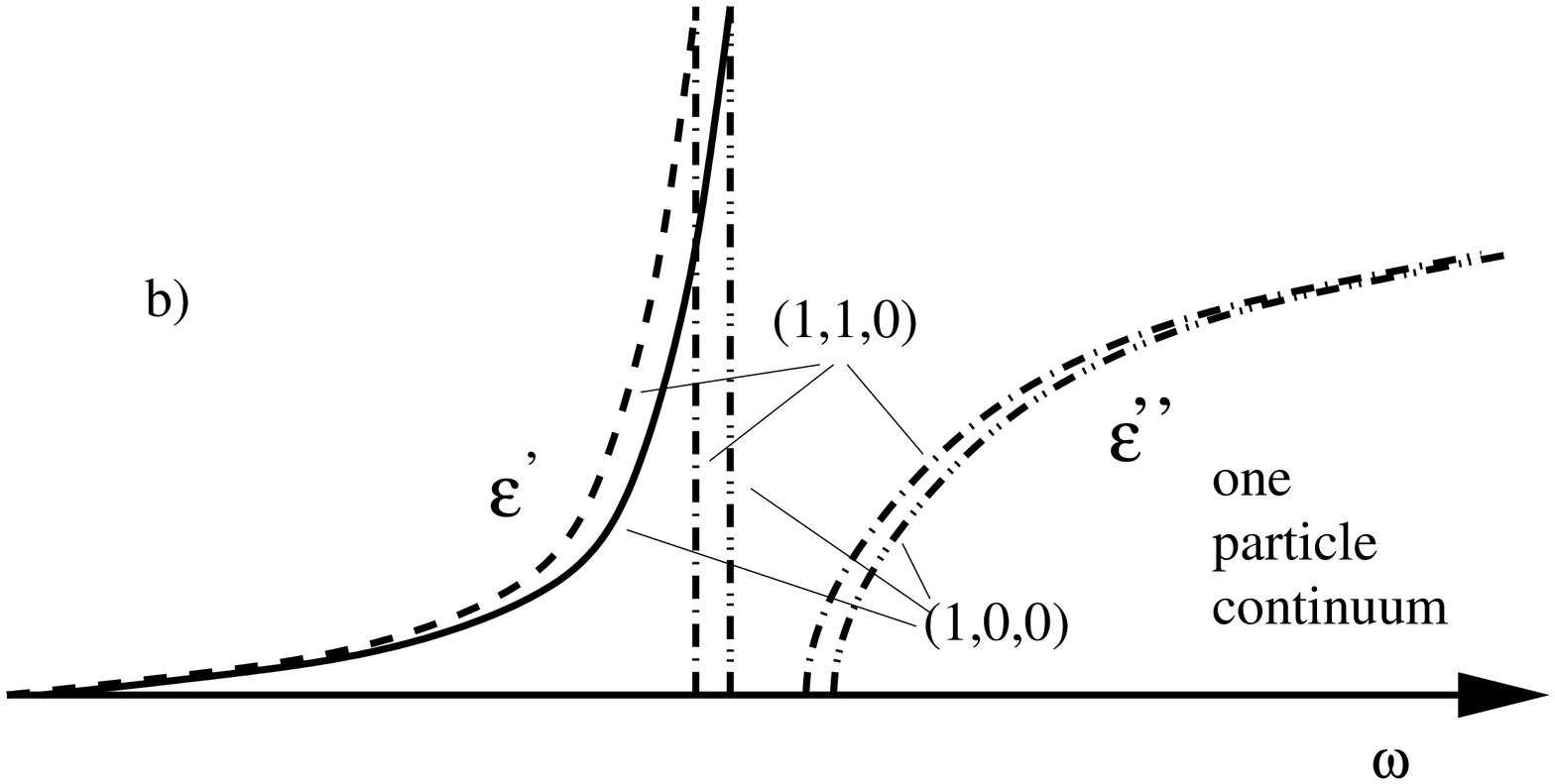}
   \end{tabular}
   \end{center}
   \caption[example] 
   { \label{fig:fig2} 
Schematic plot of the pole structure 
propper polarization $\Pi^*({\bf q},\omega)$ (proportional to 
$\epsilon_{e^--h^+}({\bf q},\omega)$)
as it can occur for different
   directions of the wave vector in an ionic crystal where a strong
   excitonic bound state is present. 
In Fig. a) we have plotted the band
   gap. The band gap can have a different shape when looking along
   different ${\bf q}$ directions. Above the band gap the two particle
   continuum starts where there is always a non zero imaginary part of
   the dielectric function. The bound state can occur below the one
   particle continuum. For simplicity we have only plotted the ground
   state level of only one exciton. The light dispersion $\omega=c q$ is
   plotted schematically. In Fig. b) the real and imaginary part of
   the dielectric function resulting from Kramers-Kronig relations
   are plotted in a schematic way. When approaching the bound state the
   real part of the
   dielectric function diverges as $(\omega - \omega_0)^{-1}$ where
   $\omega_0$ is the bound state level.
}
   \end{figure} 

\section{Analytic properties of the dielectric function}
\label{section4}

From the analyticity of $\Pi^0({\bf q},\omega)$ we can derive the
qualitative behaviour of its real and imaginary part. This is plotted
schematically in Fig. \ref{fig:fig1}. In    
Fig. \ref{fig:fig1}a the structure is shown in the $\omega - q$
plane. There is the large band gap. An excitation over the bandgap is
the creation of an electron-hole pair. Above the band gap there is a
continuum of one particle excitations. The light dispersion is shown
as well. Since the speed of light is much larger than any electronic
speed one can usually in the optical frequency range safely neglect
the wave-vector dependence of the light and use ${\bf q}=0$. This
breaks down when excitations are considered which come close to the
band gap. In this case it can become important that the band curvature
is different along different directions in the crystal. Schematically
we have plotted here a ficticious $(1,1,0)$ and $(1,0,0)$ band
curvature.

In Fig. \ref{fig:fig1}b the resulting real and imaginary part of
$\Pi^0({\bf q},\omega)$ is plotted resulting from the Kramers-Kronig
relations. In a three dimensional system the imaginary part of the one
particle continuum shows a $\sqrt{\omega-\omega_0}$
behaviour. Therefore the real part does not diverge at the band edge
(opposite to a 1D or 2D system) but increases up to a finite value.
In order to keep the plots in Fig. \ref{fig:fig1} as simple as
possible we have not plotted the real part of the polarizability above
the gap.

If for different wave-vector directions in the crystal the band gap is
shifted one can observe a spatial dispersion. This is the case in
semiconductors where no deep excitonic bound state exists.\cite{cardona71}   
Here the spatial dispersion is caused by the wave-vector dependence of
the one particle continuum. A weak attractive electron-hole
interaction will not change the picture as long as no strong excitonic
bound states are created below the one particle continuum. Therefore
in these cases the dielectric function can already be described using
$\epsilon \sim \Pi_0$. The situation is completely different in a
strongly ionic crystal like CaF$_2$ here the excitonic bound state
becomes important.

\section{The two particle bound state of the exciton}
\label{section5}

In a strongly ionic crystal screening is less effective and 
the coulomb attraction between the positively charged hole and the
negatively charged electron leads to the formation of a strong two
particle bound state. 
The solution of this two body problem involves in general a solution
of the Bethe-Salpeter equation (see e.g. \cite{rohlfing98}). However
the basic principal of this bound state formation can already be
understood with a much simpler interaction. 
The full solution for a complicated electron hole interaction for
CaF$_2$ has been
taken into account in \cite{benedict99} here we just want to highlight
the basic principles under which such a solution leads to an excitonic bound state.  
When we assume the attractive
interaction to be only a function of the total momentum of the
electron-hole pair $U({\bf q})$, the equations can be solved easily
and one gets a 
RPA (random phase approximation) denominator for the proper
polarization
which stems from a summation of a geometrical series.
\begin{equation}
\label{eq:rpa}
\Pi^*({\bf q},\omega) = 
\frac{\Pi^0({\bf q},\omega)}{1-U({\bf q})\,\Pi^0({\bf q},\omega)}
\end{equation}
This is also known as the ladder approximation to the Bethe-Salpeter
equation. New physics will occur when the denominator of Eq. (\ref{eq:rpa})
shows additional poles beside the one particle continuum. This is the
case when the denominator of Eq. (\ref{eq:rpa}) gets zero.
\begin{equation}
1-U({\bf q})\,\Pi^0({\bf q},\omega) = 0
\end{equation}
If this happens outside the one particle continuum a true bound state,
the exciton is formed. 
Finding the bound state level can be schematically seen in
Fig. \ref{fig:fig1}b. When drawing a line at $f(\omega)=1/U({\bf q})$
it can for large $U$, for large attractive interactions cross the real
part of $\Pi^0$. This crossing marks the bound state level where a new
pole occurs.
Such a situation with a bound state is schematically sketched in
Fig. \ref{fig:fig2}. Below the one particle continuum the exciton level
occurs as a true pole. The dynamic polarization and therefore also the
dielectric function has to follow Kramers-Kronig relations connecting
its real and imaginary part. In the vicinity of a true pole the
real part of the
dielectric function has to follow the asymptotic behaviour as
\begin{equation}
\label{eq:el}
\epsilon'_{e^--h^+} ({\bf q},\omega) \sim \Re\left (\Pi^*({\bf
q},\omega)\right ) \sim
\frac{1}{\omega - \omega_{ex}({\bf q})} =
\frac{2 \pi c}{\frac{1}{\lambda}-\frac{1}{\lambda_{ex}({\bf q})}}=
 \frac{2 \pi c \lambda \lambda_{ex}({\bf q})}{\lambda_{ex}({\bf q})-\lambda}
\sim \frac{1}{\lambda - \lambda_{ex}({\bf q})}
\end{equation}
The wave-vector dependence of the exciton level $\omega_{ex}(\bf q)$ (or
$\lambda_{ex}({\bf q})$) will in general differ along different directions in the
crystal. This is schematically plotted in
Fig. \ref{fig:fig2}a. Exactly this difference causes the spatial
dispersion seen in ionic crystals like CaF$_2$ and BaF$_2$. 

\section{Spatial dispersion in CaF$_2$ and predictions for
experimental observations}
\label{section6}

   \begin{figure}
   \begin{center}
   \begin{tabular}{c}
\\[-2cm]
   \includegraphics[height=11 cm]{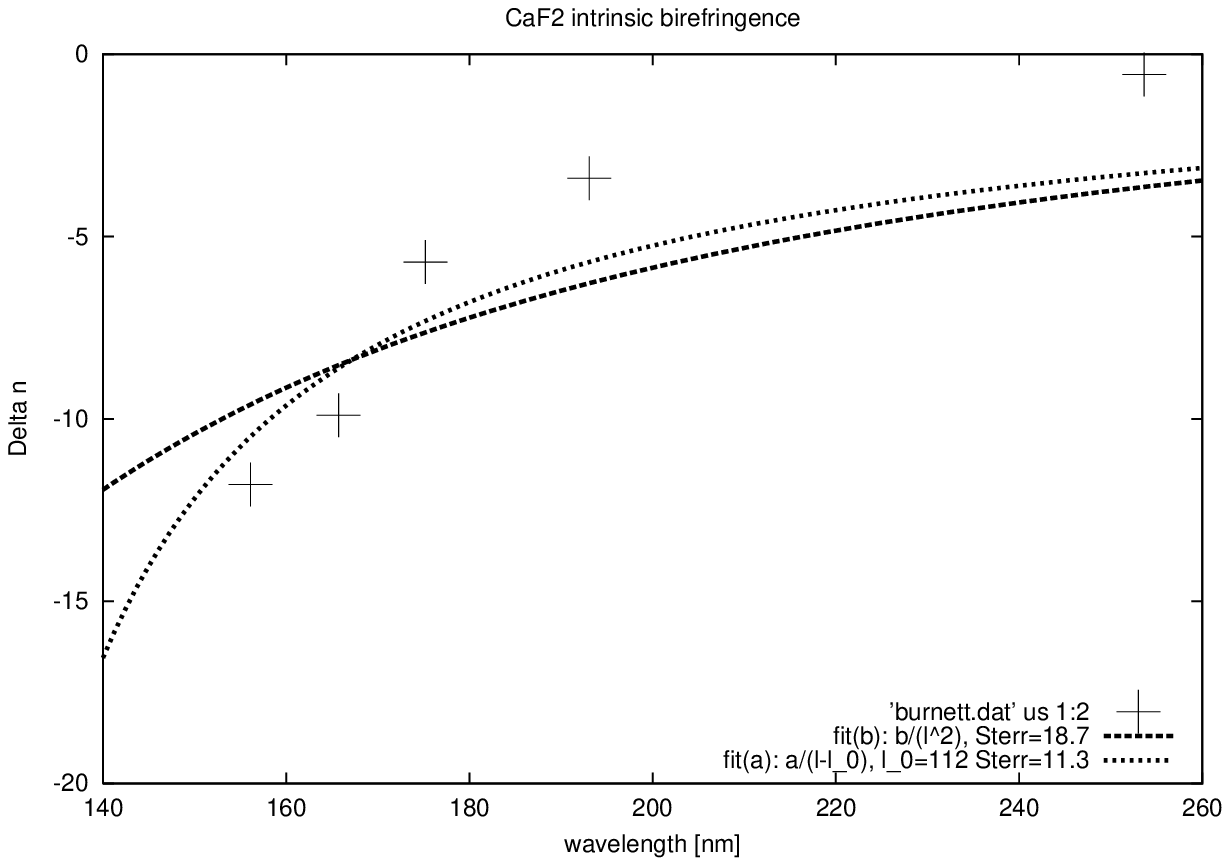}
   \end{tabular}
   \end{center}
   \caption[example] 
   { \label{fig:fig3} 
   Comparison of the data from \cite{burnett2001} (crosses) 
with a $1/\lambda^2$ (dashed line) and a
   $1/(\lambda-\lambda_{ex}({\bf q}))$ (dotted line) behaviour.
The
   $1/\lambda^2$ behaviour results from a Taylor expansion of the
   dielectric function (dashed line). A comparison with the function
   which results from the vicinity of a dispersive bound state
   (exciton) is plotted (dotted line). For the exciton wavelength the 112nm known
   from literature \cite{tomiki69} were used. The exciton
   model produces a better description of the data at smaller
   wavelength. It further makes the prediction that the dielectric
   properties will for wavelength below 157nm diverge as
   $1/(\lambda-\lambda_{ex}({\bf q}))$ when 
   aproaching the excitonic bound state.
}
   \end{figure} 

In view of the measurements performed by Burnett et
al. \cite{burnett2001} we need the refractive index along different
directions. In an isotropic medium this is where the dielectric
function is diagonal with three identical diagonal elements this would
be simply given by:
\begin{equation}
\label{eq:nsimp}
n(\omega) = \sqrt{\epsilon(\omega)}
\end{equation}
with $\epsilon$ being the (real part of the) diagonal part of the
diagonalized dielectric function. 

The concept of a refractive index becomes by far more sophisticated as
soon as the medium shows anisotropies. This is also the case for
CaF$_2$. Here one has to go into a new coordinate system whose
z-direction is determined by the ${\bf q}$-direction of the incident
light. Two possible directions perpendicular to the z-direction
complete this new coordinate system and are denoted with Greek letters
$\alpha,\beta$. Within the $\alpha$-$\beta$-plane lies the polarization
vector ${\bf e}$ pointing in the direction of the electric induction
${\bf D}$ of the light. Within this new coordinate system the
calculation of the refractive index is given by the zero of a two
dimensional determinant.\cite{landaulifshitz8}
\begin{equation}
\label{eq:n0}
\mbox{Det} \left | n^{-2} \delta_{\alpha,\beta} -
\epsilon^{-1}_{\alpha,\beta}({\bf q},\omega) \right | = 0
\end{equation}
In general the eigenvalue problem of Eq. (\ref{eq:n0}) has two
solutions which correspond to the main polarization axes.

There is a further difference to an anisotropic medium. This is the fact
that the wave-vector ${\bf q}$ is not in general parallel to the
Poynting-vector ${\bf S}$ which points along the direction of the
group velocity which is the analogon of the path in classical
geometrical optic. 
This is expected to be small in a cubic crystal.

So the refractive index is dependent on ${\bf q}$ and can for a
given experimental setup be dependent on the polarization direction of
the incident light.
\begin{equation}
n=n({\bf q},\omega) =n({\bf q},2 \pi c/\lambda) 
\end{equation}
In a cubic crystal like CaF$_2$ there are several exceptions for the
wave-vector ${\bf q}$ and the Poynting vector pointing along the same
direction where no dependence on the polarization direction
occurs as one can
derive from symmetry considerations. One is for example light
propagating along the main axis of the crystal (e.g. ${\bf q}=(1,0,0)$
now measured in the coordinate system of the crystal axes).

The dielectric function diverges close to the excitonic bound
state depending on the wave-vector ${\bf q}$ according to
Eq. (\ref{eq:el}) with
$\lambda_{ex}({\bf q})$ is the wavelength of the exciton 
for the given ${\bf q}$ vector. This wavelength must be
for CaF$_2$ close to the 112nm where the exciton is found
experimentally \cite{tomiki69}.

Note that Eq. (\ref{eq:el}) is quite different to what would result from just a
formal Taylor expansion of the dielectric function. In this case the
dielectric function should show a $1/\lambda^2$ behaviour.

We want to look at the quantity measured in \cite{burnett2001}. There
the difference of the refractive index $\Delta n$ with a wave-vector pointing in
the $(1,1,0)$ direction and the polarization vector ${\bf e}$ pointing in 
two different directions (in $(-1,1,0)$ and $(0,0,1)$) is
measured. 
Using Eq. (\ref{eq:nsimp}) gives
\begin{eqnarray}
2 n \Delta n &\approx& \Delta \epsilon\\
\Delta n & \approx & \frac{\Delta \epsilon}{2 n} \sim \frac{1}{\lambda
- \lambda_{ex({\bf q})}} 
\label{eq:dn}
\end{eqnarray}
In Fig. \ref{fig:fig3} we have plotted the wavelength dependence of
$\Delta n$ as it would result from the experimentally observed bound
state level lying  at 112nm.\cite{tomiki69}

It can be seen that describing the experimental observations with
Eq. (\ref{eq:dn}) (dotted line) gives a better agreement with the
experimental observation (crosses) than just a $1/\lambda^2$ behaviour
(dashed line) 
as it would be expected from a Taylor expansion of the
dielectric function. We therefore claim in this work that the
anisotropy which is seen in CaF$_2$ at low wavelengths is caused by
the exciton dispersion of an excitonic level around 112nm (11.2eV). 
Especially measurements in the wavelength range between 112nm and
157nm we predict to show a clear $1/(\lambda - \lambda_{ex}({\bf q}))$
behaviour. 

\section{conclusion}
\label{conclusion}

We argue  
that the optical anisotropy 
observed in CaF$_2$ is caused by exciton dispersion. The
delocalized exciton lies at approximately 112nm. Using this excitonic
model the measured data on the optical anisotropy can be well explained and
it is predicted that the dielectric function should diverge for lower
wavelengths as $1/(\lambda - \lambda_{ex}({\bf q}))$. Here $\lambda_{ex}({\bf q})$ 
is the wavelength of the excitonic bound state.

If the connection to the exciton is indeed the dominant mechanism
beyond the optical anisotropy all possible compensation approaches have
to be reviewed. E.g. forming solid solution crystals, if possible at all, would
only lead to an improvement, if the exciton dispersion will be smaller
in such a crystal. Further the subtle difference between the spatial
dispersion and birefringence has to be taken into account in a very
careful way when performing lens design. 

\acknowledgements 
M.L. thanks D.Strauch and K.Schmalzl for helpful discussions and for
band structure data on CaF$_2$. The work was supported by the BMBF
project ``Laserbasierte Ultrapr\"azisionstechnik -- 157nm Lithographie,
Teilvorhaben: Optische Materialien und Komponenten f\"ur die 157nm 
Lithographie, AP 4210''. 

%\input{appendix}

%\begin{appendix}
%\end{appendix}

%%%%%%%%%%%%%%%%%%%%%%%%%%%%%%%%%%%%%%%%%%%%%%%%%%%%%%%%%%%%%
%%%%% References %%%%%

%\bibliography{litcaf2}   %>>>> bibliography data in report.bib
%\bibliographystyle{spiebib}   %>>>> makes bibtex use spiebib.bst

\end{document}